\documentclass{aa}
\usepackage{graphics}
\begin{document}
   \thesaurus{06         
              (03.13.2;  
	       03.20.4;  
               10.08.1;  
               11.13.1;  
               12.04.1;  
               12.07.1)} 
\title{AGAPEROS: Searching for microlensing in the LMC with the Pixel
Method} \subtitle{I. Data treatment and pixel light curves production}

\author{A.-L. Melchior\inst{1,2,3}, C. Afonso\inst{4},
R. Ansari\inst{5}, {\'E.}  Aubourg\inst{4}, P. Baillon\inst{6},
P. Bareyre\inst{4}, F. Bauer\inst{4}, J.-Ph. Beaulieu\inst{7},
A. Bouquet\inst{2}, S.  Brehin\inst{4}, F.  Cavalier\inst{5},
S. Char\inst{8}, F. Couchot\inst{5}, C. Coutures\inst{4},
R. Ferlet\inst{7}, J.  Fernandez\inst{8}, C. Gaucherel\inst{4},
Y. Giraud-H\'eraud\inst{2}, J.-F. Glicenstein\inst{4},
B. Goldman\inst{4}, P. Gondolo\inst{2,9}, M. Gros\inst{4},
J. Guibert\inst{10}, D. Hardin\inst{4},
J. Kaplan\inst{2}, J. de Kat\inst{4}, M. Lachi\`eze-Rey\inst{4},
B. Laurent\inst{4}, {\'E}. Lesquoy\inst{4}, Ch. Magneville\inst{4},
B. Mansoux\inst{5}, J.-B. Marquette\inst{7}, E. Maurice\inst{11},
A. Milsztajn\inst{4}, M. Moniez\inst{5}, O. Moreau\inst{10},
L. Moscoso\inst{4}, N.  Palanque-Delabrouille\inst{4},
O. Perdereau\inst{5}, L. Pr\'ev\^ot\inst{11}, C. Renault\inst{4},
F. Queinnec\inst{4}, J. Rich\inst{4}, M. Spiro\inst{4},
A. Vidal-Madjar\inst{7}, L. Vigroux\inst{4}, S. Zylberajch\inst{4}}

\offprints{A.L.Melchior@qmw.ac.uk}

\institute{Astronomy Unit, Queen Mary and Westfield College, Mile End Road,
London E1 4NS, UK
\and Laboratoire de Physique Corpusculaire et Cosmologie (UMR 7535),
Coll\`ege de France, 75231 Paris Cedex 05, France
\and NASA/Fermilab Astrophysics Center, Fermi National Accelerator
Laboratory, Batavia, IL 60510-0500, USA
\and CEA, DSM, DAPNIA, Centre d'\'Etudes de Saclay, 911191
Gif-sur-Yvette Cedex, France
\and Laboratoire de l'Acc\'el\'erateur Lin\'eaire, IN2P3 CNRS,
Universit\'e Paris-Sud, 91405 Orsay Cedex, France
\and CERN, 1211 Gen\`eve 23, Switzerland
\and Institut d'Astrophysique de Paris, CNRS, 98 bis Boulevard Arago,
75014 Paris, France
\and Universidad de la Serena, Faculdad de Ciencias, Departemento de
Fisica, Casilla 554, La Serena, Chile.
\and Max-Planck-Institut f\"ur Physik, F\"ohringer Ring 6, 80805
M\"unchen, Germany
\and Centre d'Analyse des Images de l'INSU, Observatoire de Paris, 61
avenue de l'Observatoire, 75014 Paris, France
\and Observatoire de Marseille, 2 place Le Verrier, 13248 Marseille
Cedex 04, France}

\date{Received / Accepted}
\titlerunning{Pixel microlensing in the LMC. I}
\authorrunning{A.-L. Melchior et al.}
\maketitle

\begin{abstract} 
Recent surveys monitoring millions of light curves of resolved stars
in the LMC have discovered several microlensing events. Unresolved
stars could however significantly contribute to the microlensing rate
towards the LMC. Monitoring pixels, as opposed to individual stars,
should be able to detect stellar variability as a variation of the
pixel flux.  We present a first application of this new type of
analysis (Pixel Method) to the LMC Bar. We describe the complete
procedure applied to the EROS 91-92 data (one tenth of the existing
CCD data set) in order to monitor pixel fluxes.  First, geometric and
photometric alignments are applied to each images. Averaging the
images of each night reduces significantly the noise level. Second,
one light curve for each of the $2.1 \times 10^6$ pixels is built and
pixels are lumped into 3.6$\arcsec \times$ 3.6$\arcsec$ super-pixels,
one for each elementary pixel. An empirical correction is then applied
to account for seeing variations. We find that the final super-pixel
light curves fluctuate at a level of 1.8\% of the flux in blue and
1.3\% in red. We show that this noise level corresponds to about twice
the expected photon noise and confirms previous assumptions used for
the estimation of the contribution of unresolved stars. We also
demonstrate our ability to correct very efficiently for seeing
variations affecting each pixel flux. The technical results emphasised
here show the efficacy of the Pixel Method and allow us to study
luminosity variations due to possible microlensing events and variable
stars in two companion papers.

\end{abstract}

\section{Introduction} 
The amount and nature of Dark Matter present in the Universe is an
important question for cosmology (see e.g. White et al.
(1996)\nocite{White:1996} for current status). On galactic scales
(Ashman 1992)\nocite{Ashman:1992}, dynamical studies (Zaritsky
1992)\nocite{Zaritsky:1992} as well as macrolensing analyses (Carollo
et al.  1995)\nocite{Carollo:1995} show that up to 90\% of the
galactic masses might not be visible. One plausible explanation is
that the stellar content of galaxies is embedded in a dark halo.
Primordial nucleosynthesis (Walker et al. 1991; Copi et al. 1995)
\nocite{Walker:1991,Copi:1995} predicts a larger number of baryons
than what is seen (Persic \& Salucci 1992)\nocite{Persic:1992}, and so
dark baryons hidden in gaseous or compact objects (Carr 1994, Gerhard
\& Silk 1996)\nocite{Carr:1994,Gerhard:1996} could explain, at least
in part, the dark galactic haloes.

In 1986, Paczy\'nski\nocite{Paczynski:1986} proposed microlensing
techniques for measuring the abundance of compact objects in galactic
haloes. The LMC stars are favourable targets for microlensing events
searches. Since 1990 and 1992, the EROS (Aubourg et
al. 1993)\nocite{Aubourg:1993} and MACHO (Alcock et
al. 1993)\nocite{Alcock:1993} groups have studied this line of
sight. The detection of 10 microlensing events has been claimed in the
large mass range $0.05 - 1 M_{\odot}$ (Aubourg et al. 1993, Alcock et
al. 1996)\nocite{Aubourg:1993,Alcock:1996}. This detection rate,
smaller than expected with a full halo, indicates that the most likely
fraction of compact objects in the dark halo is $f = 0.5$ (Alcock et
al. 1996)\nocite{Alcock:1996}. Concurrently, the small mass range has
been excluded for a wide range of galactic models by the EROS and
MACHO groups.  Objects in the mass range ($5\times 10^{-7} M_\odot < M
< 5 \times 10^{-4} M_\odot$) could not account for more than 20\% of
the standard halo mass (Alcock et al. 1998)\nocite{Alcock:1998}.  In
the meantime, the DUO (Alard et al. 1995)\nocite{Alard:1995}, MACHO
(Alcock et al.  1995)\nocite{Alcock:1995a} and OGLE (Udalski et al.
1995)\nocite{Udalski:1995b} groups look towards the galactic bulge
where star-star events are expected. The detection rate is higher than
expected from galactic models (see for instance Evans 1994, Alcock et
al. 1995, Stanek et
al. 1997)\nocite{Evans:1994,Alcock:1995a,Stanek:1997}.  The events
detected in these two directions demonstrate the efficacy of the
microlensing techniques based on the monitoring of several millions of
stars.

\subsection*{Microlensing Searches with the Pixel Method}
\nocite{Ansari:1997a} The detection of a larger number of events is
one of the big challenges in microlensing searches. This basically
requires the monitoring of a larger number of stars. The Pixel Method,
initially presented by Baillon et al. (1993)\nocite{Baillon:1993},
gives a new answer to this problem: monitoring pixel fluxes. On images
of galaxies, most of the pixel fluxes come from unresolved stars,
which contribute to the background flux. If one of these stars is
magnified by microlensing, the pixel flux will vary
proportionally. Such a luminosity variation can be detected above a
given threshold, provided the magnification is large enough.  Unlike
other approaches (namely star monitoring and Differential Image
Photometry, see below), the Pixel Method {\em does not} perform a
photometry of the stars but is designed to achieve a high efficiency
for the detection of luminosity variations affecting unresolved
stars. This means that we will work with pixel fluxes and {\em not}
with star fluxes.  A theoretical study of the pixel lensing method has
been published by Gould (1996b)\nocite{Gould:1996}.

This pixel monitoring approach has two types of application. Firstly,
it allows us to investigate more distant galaxies and thus to study
other lines of sight.  This has led to observations of the M31
galaxy. The AGAPE team (Ansari et al. 1997)\nocite{Ansari:1997a} has
shown that this method works on M31 data, and luminosity variations
compatible with the expected microlensing events have been detected
but the complete analysis is still in progress (Giraud-H\'eraud
1997)\nocite{YGH:1997}.  A similar approach, though technically
different, called Differential Image Photometry is also investigated
by the VATT/Columbia collaboration (Crotts 1992, Tomaney \& Crotts
1996)\nocite{Crotts:1992,Tomaney:1996}. Some prospective work has also
been done towards M87 (Gould 1995)\nocite{Gould:1995b}.

The second possibility is to apply pixel microlensing on existing
data, thus extending the sensitivity of previous analyses to
unresolved stars.  This is precisely the subject of this paper and of
the two which will follow: we present the implementation of the Pixel
Method on CCD images of the LMC.

\subsection*{Pixel Method on the LMC}
We have applied for the first time a comprehensive pixel analysis on
existing LMC images collected by the EROS collaboration. With respect
to previous analyses (Queinnec 1994, Aubourg et al. 1995, Renault
1996)\nocite{Queinnec:1994,Aubourg:1995,Renault:1996}, our analysis of
the same data using pixel monitoring allows us to extend the mass
range of interest up to $1\, M_\odot$ and to increase the sensitivity
of microlensing searches. On these images, a large fraction of the
stars remains unresolved: typically 5 to 10 stars contribute to 95\%
of the pixel flux in one square arc-second.  Since this approach
potentially uses all the image content (and not only the resolved
stars), the volume of the data to handle is much larger. Hence we
perform this first exploratory analysis on a relatively small data
set: 0.25 deg$^2$ covering a period of observation of 120 days, which
corresponds to 10\% of the LMC CCD data (91-94).

This paper is the first of a series of three, describing the data
treatment (this paper), the microlensing search (Melchior et
al. 1998a, hereafter Paper II) and a catalogue of variable stars
(Melchior et al. 1998b, hereafter Paper III). In the companion papers
(Papers II and III), we show how the data treatment described here to
produce pixel light curves allows us to perform analyses that increase
the sensitivity to microlensing events and variable stars with respect
to the star monitoring analysis applied on the same field: an order of
magnitude in the number of detectable luminosity variations is gained.

To discover real variations, the images and light curves have to be
corrected for various sources of fake variabilities, such as
geometrical and photometric mismatch, or seeing changes between
successive images. The construction of light curves cleaned from these
effects is the subject of this first paper. If the flux of a given
star contributes to the pixel flux, the latter can be expressed as
follows:
\begin{equation}
\phi_{pixel}   = f \times \phi_{\rm star} + \phi_{bg} ,
\label{eq:phipix}
\end{equation}
where $\phi_{\rm star}$ is the flux of the given star, $f$ the
fraction of the star flux that enters the pixel, hereafter called
seeing fraction and $\phi_{bg}$ corresponds to the flux of all other
contributing stars plus the sky background.

If this particular star exhibits a luminosity variation, then we will
be able to detect it as a variation of the pixel flux:
\begin{equation}
\Delta \phi_{pixel}   = f \times \Delta \phi_{\rm star} ,
\label{eq:delphipix}
\end{equation}
provided it stands well above the noise. Actually, this pixel flux is
affected by the variations of the observational conditions and our
goal here is to correct for them. We discuss the level of noise
achieved after these corrections and include this residual noise in
error bars.  The outline of this paper is as follows.  In
Sect. \ref{section:data}, we start with a short description of the
data used. In Sect. \ref{section:align}, we successively describe the
geometric and photometric alignments applied to the images. We are
thus able to build pixel light curves and to discuss their stability
after this preliminary operation. In Sect. \ref{section:meanimage}, we
average the images of each night, thus reducing the fluctuations due
to noise considerably. In Sect. \ref{subsection:stabi}, we consider
the benefits of using super-pixel light curves. In
Sect. \ref{section:seecorr}, we correct for seeing variations and
obtain light curves cleaned from most of the changes in the
observational conditions. At this stage, a level of fluctuations
smaller than 2\% is typically achieved on the super-pixel fluxes.  In
order to account for the noise present on the light curves, we
estimate, in Sect. \ref{section:opteb}, an error for each super-pixel
flux.  We conclude in Sect. \ref{section:simu} that the light curves
of super-pixels, resulting from the complete treatment, reach the
level of stability close to the expected photon noise. They are
therefore ready to be used to search for microlensing events and
variable objects, as presented in the companion Papers II and III.

\section{The data}
\label{section:data}
\subsection{Description of the data set}
The data have been collected at La Silla ESO in Chile with a $40$cm
telescope ($f/10$) equipped with a thick CCD camera composed of
$8\times2$ CCD chips of $400\times579$ pixels with scale of
$1.21\arcsec$/pixel (Arnaud et al., 1994b\nocite{Arnaud:1994b},
Queinnec, 1994\nocite{Aubourg:1995,Queinnec:1994} and Aubourg et al.,
1995).  The gain of the camera was $6.8 {e^{-}}/$ADU with a read-out
noise of 12 photo-electrons.  For the 1991-92 campaign only 11 chips
out of 16 were active. Due to technical problems, we only analyse 10
of them. The monitoring has been performed in two wide colour bands
(Arnaud et al., 1994a\nocite{Arnaud:1994a}). Exposure times were set
to 8 min in red ($\langle{\lambda}\rangle = 670$ nm) and 15 min in
blue ($\langle{\lambda}\rangle = 490$ nm). As the initial goal was to
study microlensing events with a short-time scale (Aubourg et
al. 1995)\nocite{Aubourg:1995}, up to 20 images per night in both
colours are available.  A total of 2000 blue and red images were
collected during 95 nights spread over a 120 days period (18 December
1991 - 11 April 1992).  The combined CCD and filter efficiency curves
as shown in Grison et al. (1995)\nocite{Grison:1994b} lie below 15\% in
blue and below 35\% in red.  Bias subtraction and flat-fielding have
been performed on-line by the EROS group.

The seeing varies between $1.6$ and $3.6$ arc-second with a mean value
of $2.9$ arc-second (typical dispersion $0.5$ arc-second).  It should
be emphasised that the observational strategy (exposure time) has been
optimised for star monitoring. In other words, this means that the
photon noise associated with the mean flux (typically 280 ADU per
pixel in red and 100 ADU in blue) is relatively large: 6.6 ADU in red
and 3.8 ADU in blue. To apply the Pixel Method to this data set, we
take advantage of the large number of images available per night,
increasing the signal-to-noise ratio with an averaging procedure.

\subsection{Absolute calibration}
The procedures described below are performed with respect to a
reference image. The correspondence between the flux measured on these
images and the magnitude, deduced from Grison et al
(1995)\nocite{Grison:1994b}, is as follows:
\begin{equation}
m_B = -2.5  \log \phi_B + 24.8
\end{equation}
\begin{equation}
m_R = -2.5  \log \phi_R + 24.9
\end{equation}
where $\phi_B$ and $\phi_R$ are the flux of a star in ADU in the blue
and red respectively.  Note that the zero point is about the same in
the two colours, whereas the background flux is much larger in red
than in blue.  The correspondence with the Johnson-Cousins system can
be found in Grison et al. (1995).

\vspace{0.5cm} The aim of the whole treatment presented below is to
obtain pixel light curves properly corrected for variations of the
observational conditions. The PEIDA package used by the EROS group was
adapted for pixel monitoring. This treatment is applied to the first
CCD campaign (1991-92) of the EROS group on the LMC bar, i.e. 10\% of
the whole data set analysed in Renault (1996)\nocite{Renault:1996}.

\section{Image alignments}
\label{section:align}
The alignments described in this section are needed in order to build
pixel light curves from images that are never taken under the same
observing conditions. Firstly, the telescope never points exactly
twice in the same direction so that the geometric alignment must
ensure that the same area of the LMC contributes to the same pixel
flux, through the entire period of observation. Secondly, photometric
conditions, atmospheric absorption and sky background light change
from one frame to another. The photometric alignment corrects for
these global variations.

Errors affecting pixel fluxes after these corrections are a key issue
as discussed through this section. It is not obvious how to
disentangle the various sources of error introduced at each step, in
particular after the geometric alignment. Global errors for each pixel
flux, including {\em all} sources of noise, will be evaluated in
Sect. \ref{section:opteb}.

\subsection{Geometric alignment}
\label{subsection:geo}
Between exposures, images are shifted by as much as 40 pixels and this
displacement has to be corrected, in order to ensure that each pixel
always covers the same area of the LMC. As emphasised below, errors
affect the pixel flux after the geometric alignment and two components
can be distinguished. The first one, resulting from the uncertainty in
the parameters of displacement, turns out to be negligible, whereas
the second one, introduced by the linear interpolation, is a more
important source of noise. In this sub-section, we give a qualitative
overview of these sources of errors. This study, based on synthetic
images, allows us to disentangle errors due to the geometrical
alignment from other effects present on real images, because the
position and content of unaligned synthetic frames are known by
construction.

\paragraph*{Displacement parameters.}\hspace{0.5cm}
The parameters of displacement are determined with the PEIDA algorithm
(Ansari, 1994\nocite{Ansari:1994}), based on the matching of star
positions. Beside translation, rotation and dilatation are also taken
into account as far as their amplitude remains small (otherwise the
corresponding images are removed from further consideration).

A series of mock images synthesised with the parameters of real images
(geometric displacement, absorption, sky background and seeing) allows
us to estimate the mean error on the pixel position to be $0.011 \pm
0.005$ pixel.  Similar estimates have been obtained by the EROS group
(Ansari, private communication) on real data.

This introduces a small mismatch between pixel fluxes: in first
approximation, the error on the flux is proportional to the pixel area
corresponding to the difference between the true and the computed
pixel position.
\paragraph*{Linear interpolation.}\hspace{0.5cm}
\label{pg:phys}
 Once the parameters of displacement are estimated, pixel fluxes are
corrected with a linear interpolation.  This interpolation is
necessary in order to monitor pixel fluxes, and to build pixel light
curves.
\begin{figure}
\resizebox{\hsize}{!}{\includegraphics{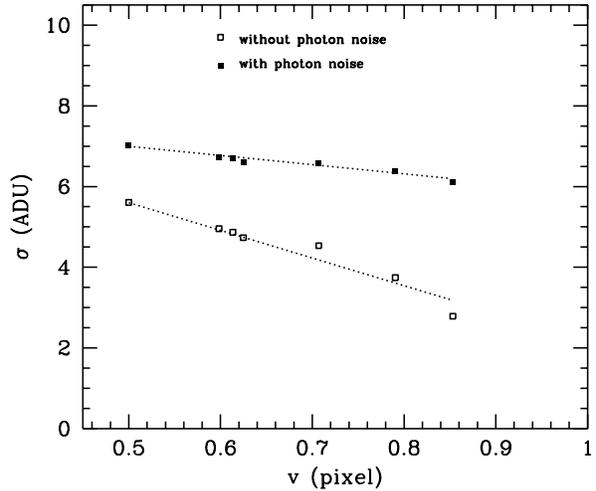}}
\caption{Error due to linear interpolation estimated with two sets of
synthetic images: \protect$\sigma$ is the dispersion measured on the
flux difference between pixels on the ``reference'' image and
corrected images, while \protect $v$ is a function of the
displacement, as discussed in the text (Eq.~\ref{eq:v}).}
\label{fig:trans4}
\end{figure}
We use synthetic images to understand qualitatively the residual
errors.  Two sets of blue images are simulated with the identical
fluxes (new moon condition) and seeings (2.5~arc-second) but shifted
with respect to one of them (the ``reference'' image). A linear
interpolation is applied to each of these images in order to match the
position of the reference.  In case of pure translation, the corrected
flux is computed with the flux of the 4 pixels overlapping the pixel
$p$ on the reference frame: the areas of these intersections with this
pixel $p$ are used to weight each pixel flux. The square of the
variable $v$, depending upon $\delta x$ and $\delta y$, the
displacement in the $x$ and $y$ directions,
\begin{equation}
v = \sqrt{\left(\delta x^2+{\left(1-\delta x\right)}^2\right)
\left(\delta y^2+{\left(1-\delta y\right)}^2\right)} ,
\label{eq:v}	
\end{equation}
is the sum of the square of these overlapping surfaces. It
characterises the mixing of pixel fluxes produced by this
interpolation: the smaller $v$ is, the more pixels are mixed by the
interpolation.

Figure \ref{fig:trans4} displays an estimate of the residual errors
affecting pixel fluxes for different displacement parameters, and
shows a correlation of the errors with the variable $v$.  The first
set of images, simulated without photon noise, shows errors on pixel
fluxes due to linear interpolation smaller than 5.5 ADU (about 4.5\%
of the mean flux). The second set of images, simulated with photon
noise, allows us to check that the photon noise adds quadratically
with the ``interpolation'' noise and that residual errors are smaller
than 7 ADU. The correlation observed on this figure between the error
$\sigma$ and the variable $v$ can be understood as follows: when $v$
decreases, the interpolated image gets more and more degraded, and the
interpolation noise increases while the poisson noise is smeared out.

This residual error is strongly seeing dependent. If the above
operation is performed on an image with a seeing of $2$ arc-second,
the residual errors are as large as 10\% of the mean flux: the larger
the seeing difference, the larger the residual error.  As the seeing
of raw images varies between $1.6$ and $3.6$ arc-second, this makes a
detailed error tracing very difficult.  The PSF is also slightly
widened due to the re-sampling, but this effect remains small compared
to other sources of PSF variability, and is largely averaged out when
summing over the images of a night (see
Sect.~\ref{section:meanimage}).

\subsection{Photometric alignment}
\label{section:phot}
\begin{figure}
\resizebox{\hsize}{!}{\includegraphics{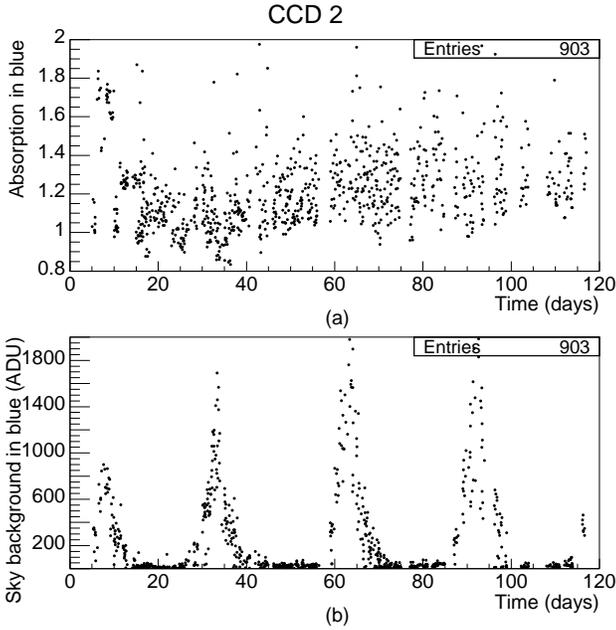}}
\caption{Photometric alignment: (a) absorption and (b) sky background
flux (in ADU) estimated in each blue image.}
\label{fig:fdc}
\end{figure}
Changes in observational conditions (atmospheric absorption and
background flux) are taken into account with a global correction
relative to the reference image. We assume that a linear correction is
sufficient:
\begin{equation}
\phi_{\rm corrected} = a \phi_{raw} + b ,
\label{eq:ap}
\end{equation}
where $\phi_{\rm corrected}$ and $\phi_{raw}$ are the pixel fluxes
after and before correction respectively.  The absorption factor $a$
is estimated for each image with a PEIDA procedure, based on the
comparison of star fluxes between this image and the reference frame
(Ansari, 1994\nocite{Ansari:1994}). A sky background excess is
supposed to affect pixel fluxes by an additional term $b$ which
differs from one image to another.

In Fig. \ref{fig:fdc}, we plot the absorption factor (top) and the sky
background (bottom) estimated for each image with respect to the
reference image as a function of time.  The absorption can vary by as
much as a factor $2$ within a single night. During full moon periods,
the background flux can be up to 20 times higher than during moonless
nights, increasing the statistical fluctuations by a factor up to
4.5. However, this high level of noise concerns very few images (see
Fig.~\ref{fig:fdc}), and only about 20\% of the images more than
double their statistical fluctuations. Despite their large noise, full
moon images improve the time sampling, and at the end of the whole
treatment, the error bars associated with these points are not
significantly larger than those corresponding to new moon periods,
except for a few nights.

\subsection{Residual large-scale variations and their correction}
\begin{figure}
\resizebox{\hsize}{!}{\includegraphics{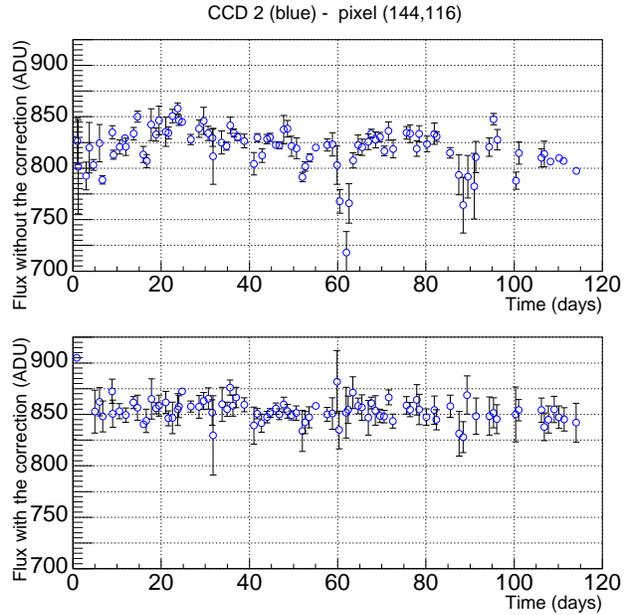}}
\caption{Pixel light curve before (above) and after (below) filtering
out the large-scale spatial variations. Fluxes are given in ADU.}
\label{fig:CLavap}
\end{figure}

We note the presence of a variable spatial pattern particularly
important during full moon periods. This residual effect, probably due
to reflected light, can be eliminated with a procedure similar to that
applied to the AGAPE data, as described by Ansari et al. (1997). We
calculate a median image with a sliding window of $9\times 9$ pixels
on the difference between each image and the reference image. It is
important to work on the difference in order to eliminate the
disturbing contributions of stars, and to get a median that retains
only large-scale spatial variations. We then subtract the
corresponding median from each image, to filter out large-scale
spatial variations.  In Fig.  \ref{fig:CLavap}, we show a light curve
before and after this correction. Above, the pixel light curve
presents important systematic effects during full moon periods,
effects which have disappeared below, after correcting for these
large-scale variations.

\subsection{Image selection}
After these alignments, we eliminate images whose parameters lie in
extreme ranges. We keep images which have no obvious defects and
parameters in the following range:
\begin{itemize}
\item Absorption factor:\\
 $0.6 < a^{R} < 1.5$ ; $0.6 < a^{B} < 1.5$
\item Mean flux (ADU):\\ $100.0 < \phi^{R} < 2000.0$\\ $70.0 <
\phi^{B} < 1500.0$
\item Seeing (arcsec):\\
 $ S^{R} < 3.6$ ; $ S^{B} < 3.6$
\end{itemize}
This procedure rejects about 33\% of the data.

\subsection{Stability of elementary pixels after alignment}
\label{sec:err}
\begin{figure}
\resizebox{\hsize}{!}{\includegraphics{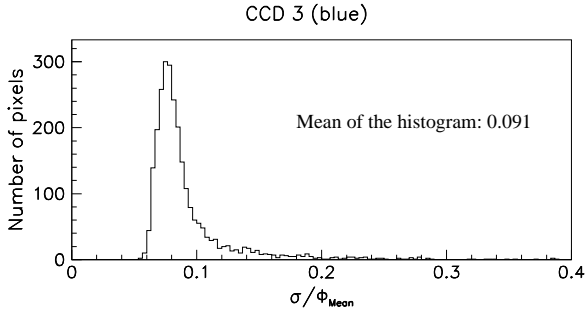}}
\caption{Relative flux stability of about 1000 flux measurements in
the blue band, spread over 120 days for {\it pixels} within a
$50\times 50$ patch of CCD~3.}
\label{fig:stab}
\end{figure}
\begin{figure}
\resizebox{\hsize}{!}{\includegraphics{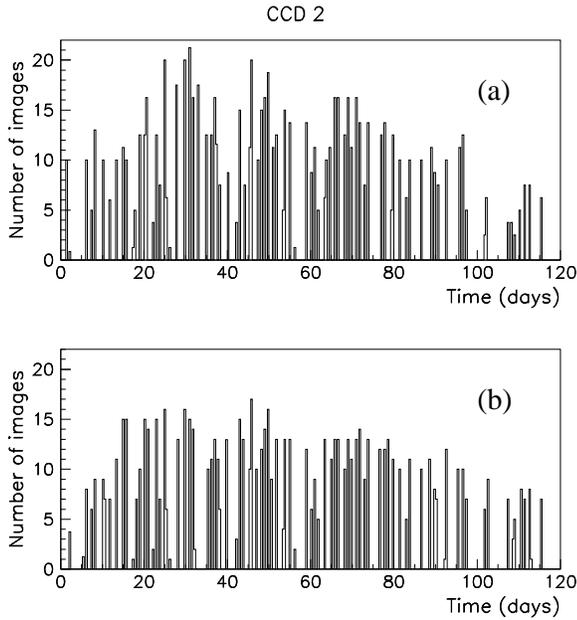}}
\caption{Number of images per night for one CCD field: in red (a) and
in blue (b).}
\label{fig:nbima}
\end{figure}
We are now able to build pixel light curves, made of about 1000
measurements spread over 120 days.  The stability can be expressed in
term of the relative dispersion ${\sigma}/{\phi}$ measured for each
light curve, where $\phi$ stands for the mean flux and $\sigma$ for
the dispersion of the light curve.  This dispersion gives us a global
estimate of the errors introduced by the alignments, combined with all
other sources of noise (photon noise, read-out noise\ldots). In
Fig. \ref{fig:stab}, we present the histogram of this dispersion for
one $50\times 50$ patch of one CCD field, which shows a mean
dispersion of 9.1\%. We estimate the contribution of the photon noise
alone to be as high as 7\%.  With such a noise level, dominated at
this stage by photon counting and flux interpolation errors, one does
not expect a good sensitivity to luminosity variations.  Fortunately,
various improvements described in the following (namely the averaging
of the images of each night, the super-pixels and the seeing
correction) will further reduce this dispersion by a factor of 5.

\section{Going to one image per night}
\label{section:meanimage}
\begin{figure}
\resizebox{\hsize}{!}{\includegraphics{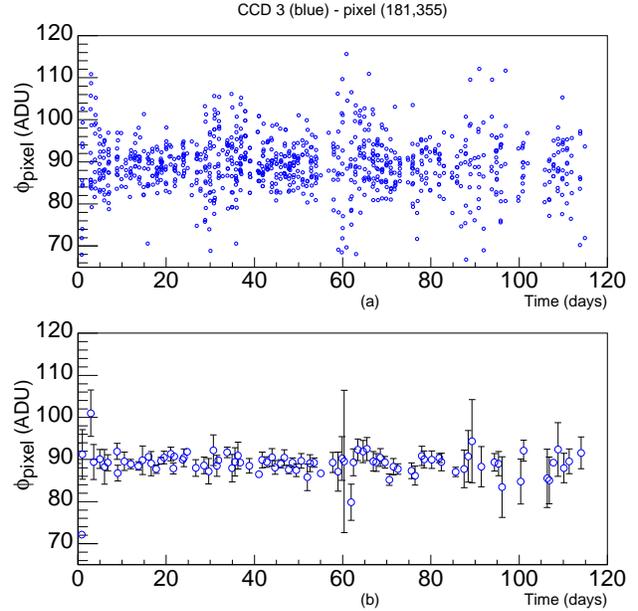}}
\caption{A stable pixel light curve (a) before and (b) after the mean
is performed over each night.}
\label{fig:cl1}
\end{figure}
\begin{figure}
\resizebox{\hsize}{!}{\includegraphics{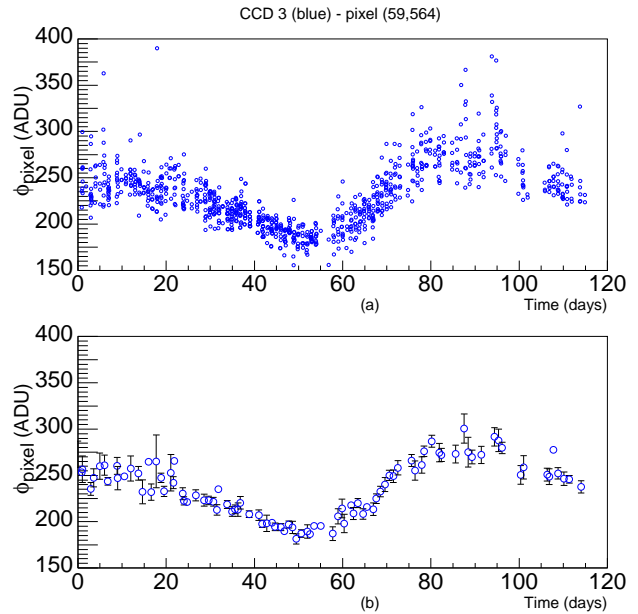}}
\caption{A variable pixel light curve before (a) and after (b) the
mean is performed over each night.}
\label{fig:cl2}
\end{figure}
The motivation of this pixel analysis is to increase the sensitivity
to long duration events ($\ge 5$ days) in the mass range where all the
known candidates have been observed. It is crucial to note that a
sampling rate of 1 measurement per day is sufficient. The numerous
images available each night (up to 20 per night) allow us to reduce
the noise discussed in Sect. \ref{sec:err}, by co-adding them, and are
very useful for the error estimation as emphasised in
Sect. \ref{section:opteb}.

\subsection{Construction}
\label{subsection:cons}
We average the images of each night.  During the night $n$, we have,
for each pixel $p$, $N^p_n$ measurements of flux ($ \phi^p_{n,j} $; $
j = 1,N^p_n $). The number of measurements $N^p_n$ available each
night is shown in Fig. \ref{fig:nbima} and ranges between 1 and 20
with an average of 10.  The mean flux $\phi^p_n$ of pixel $p$ over the
night is computed removing the fluxes which deviate by more than
$3\sigma$ from the mean, in order to eliminate any large fluctuation
due to cosmic rays, as well as CCD defects and border effects. Note
that, due to this cut-off, the number of measurements $N^p_n$ used for
a given night can differ from pixel to pixel.
\begin{figure}
\resizebox{\hsize}{!}{\includegraphics{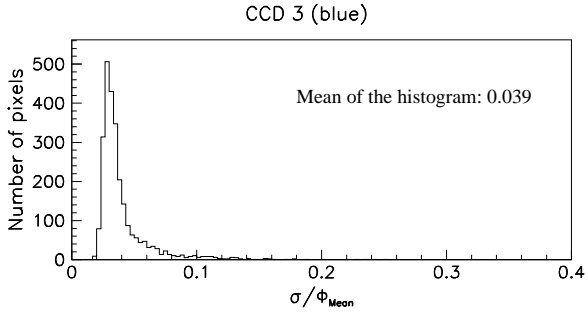}}
\caption{Relative flux stability achieved on {\it pixel} light curves
after averaging the images of each night, on a $50\times 50$ patch of
CCD~3.}
\label{fig:stabpm}
\end{figure}

\subsection{Results}
Figure \ref{fig:cl1} shows the result of this operation on a typical
pixel light curve. The dispersion in the data on the top panel (a) is
reduced and included in the error bars (see in
Sect. \ref{section:opteb}) as shown on the bottom panel (b). Figure
\ref{fig:cl2} shows the same operation applied to a pixel light curve
exhibiting a long time scale variation. One can notice that
uncertainties in the data during full moon periods are not
systematically larger than those corresponding to new moon periods.
Figure \ref{fig:stabpm} displays the histogram of relative stability
for the resulting light curves, for the same area as for
Fig. \ref{fig:stab}. A mean dispersion of 3.9\% is measured: the noise
is thus reduced by more than a factor 2. Photon noise is estimated to
be 3.3\%.

\subsection{Additional remarks}
Thanks to this procedure the PSF of the composite images will tend
towards a Gaussian. This thus removes the inhomogeneity in the PSF
shape that can be observed on raw images.  In particular, the seeing
on these composite images becomes more homogeneous with an average
value of $3.0$ arc-second in red and $2.9$ arc-second in blue and a
quite small dispersion of 0.25 arc-second. The seeing dispersion is
divided by a factor 2 with respect to the initial individual images,
whereas the average value is similar.

To summarise, this procedure improves the image quality, reduces the
fluctuations that could come from the alignments and removes cosmic
rays.

\section{Super-pixel light curves}
\label{subsection:stabi}
\begin{figure}
\resizebox{\hsize}{!}{\includegraphics{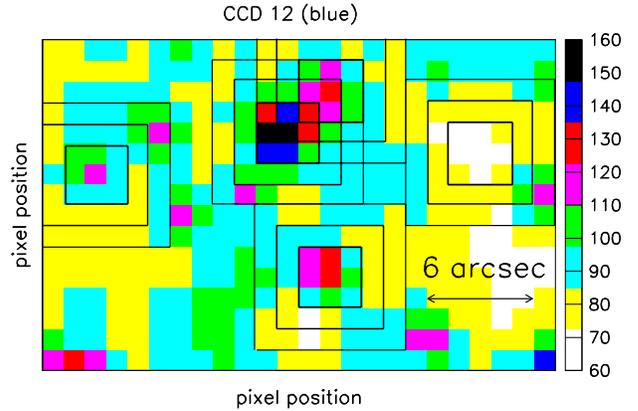}}
\caption{Example of a \protect 19''$\times$29'' field on our data in
blue, with 3 arcsec seeing. The grey scale gives the intensity in
ADU.}
\label{fig:patch}
\end{figure}
\begin{figure}
\resizebox{\hsize}{!}{\includegraphics{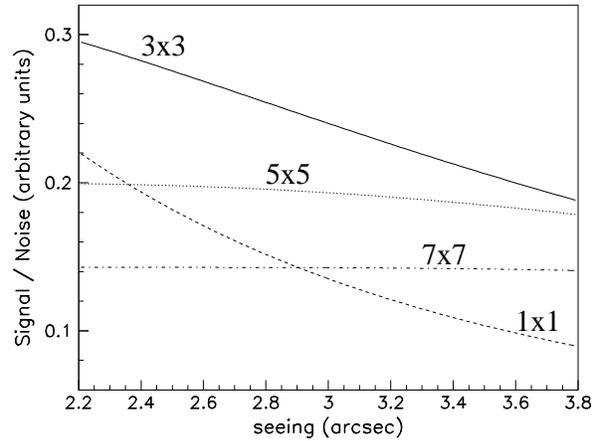}}
\caption{The signal to noise ratio expected for a single star, whose
centroid lies in the central pixel, is given as a function of the
seeing values for different super-pixel sizes. This assumes a circular
Gaussian PSF.}
\label{fig:erf}
\end{figure}
So far we have worked with elementary pixel light curves. Pixels which
cover 1.21''$\times$1.21'' are much smaller than the typical seeing
spot and receive on average only 20\% of the flux of a star, whose
centre lies in the pixel.  A significant improvement on the light
curves stability can be further achieved by considering super-pixel
light curves. Super-pixels are constructed with a running window of
$d_{sp} \times d_{sp}$ pixels, keeping as many super-pixels as there
are pixels, and their flux is the sum of the $d_{sp}^2$ pixel
fluxes. These super-pixels have to be taken large enough to encompass
most of the flux of a centred star, but not too large in order to
avoid surrounding contaminants and dilution. As such, their size
should be optimised for this dense star field given the seeing
conditions.

Figure \ref{fig:patch} illustrates the different super-pixel sizes
that can be considered.  The expected signal to noise (S/N) ratio is
proportional to the ratio of the seeing fraction to the super-pixel
size ($d_{sp}$).  Going from $1\times 1$ to $3\times 3$ super-pixels
increases the seeing fraction by more than a factor 3. Then increasing
the super-pixel size further increases the seeing fraction
substantially less than the fluctuations of the sky background.
Figure \ref{fig:erf} displays the variation with seeing of the signal
to noise ratio for different super-pixel sizes.  It is clear that
$3\times 3$ super-pixels offer the best S/N ratio for our
configuration.
\begin{figure}
\resizebox{\hsize}{!}{\includegraphics{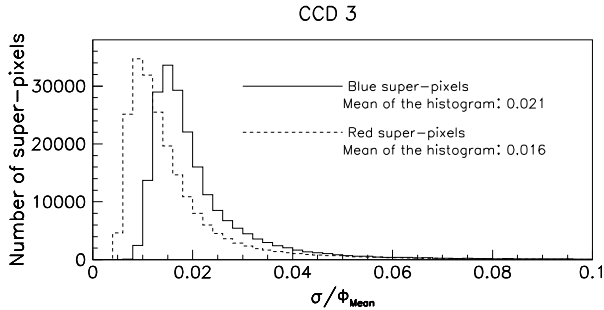}}
\caption{Flux stability achieved on {\it super-pixel} light curves in blue
(a) and in red (b) for all the pixels of CCD~3 (before the seeing
correction).}
\label{fig:stabi1}
\end{figure}

As discussed by Ansari et al (1997)\nocite{Ansari:1997a}, the
alternative that consists in taking the average of the neighbouring
pixels weighted with the PSF is not appropriate here, as it amplifies
the fluctuations due to seeing variations.

Figure \ref{fig:stabi1} shows the relative dispersion affecting the
super-pixel fluxes for CCD~3: we measure in average 2.1\% in blue and
1.6\% in red, which corresponds to about twice the estimated level of
photon noise (1.1\% in blue and 0.7\% in red). The comparison with
Fig. \ref{fig:stabpm} shows that the dispersion is reduced by a factor
smaller than $\sqrt{ 9\, {\rm (pixels)}} = 3$ because of the
correlation between neighbouring pixels. This stability can be
improved even further by correcting for seeing variations.

\section{Seeing correction}
\label{section:seecorr}
Despite of the stability discussed above, fluctuations of super-pixel
fluxes due to seeing variations are still present. For a star lying in
the central pixel (of the $3 \times 3$ super-pixel), on average 70\%
of the star flux enters on average the super-pixel for a Gaussian PSF,
but this seeing fraction is correlated with the changing seeing. In
this sub-section, we show that this correlation is linear and can be
largely corrected for.

\subsection{Correlation between flux and seeing}
\label{sect:correl}
\begin{figure}
\resizebox{\hsize}{!}{\includegraphics{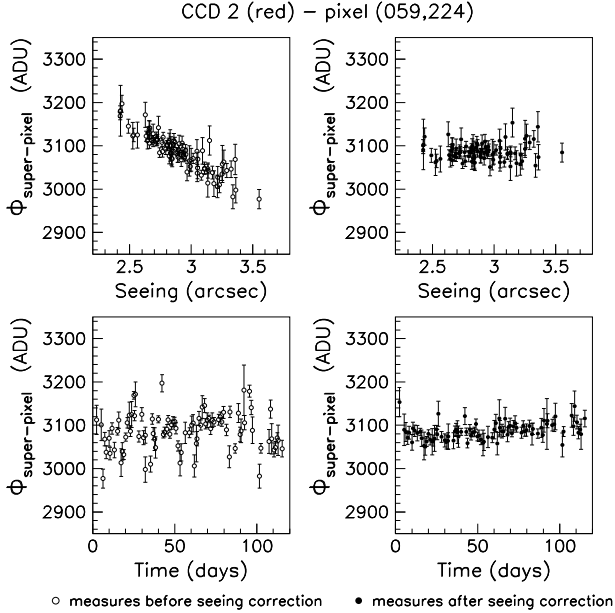}}
\caption{Seeing correction applied to the flux of one super-pixel
light curve anti-correlated with the seeing \protect ($\rho^p
=-0.89$). This corresponds to the super-pixel dominated by a centred
resolved star, whose position on the CCD frame is labelled ``A'' in
Fig. \ref{fig:pat1}. The error bars shown here and in the following
figures are computed as described in Sect. \ref{section:opteb}. }
\label{fig:clsee1}
\end{figure}

Depending on their position with respect to the nearest star,
super-pixel fluxes can significantly anti-correlate with the seeing if
the super-pixel is in the seeing spot, or correlate if instead it lies
in the tail of a star.
\begin{figure}
\resizebox{\hsize}{!}{\includegraphics{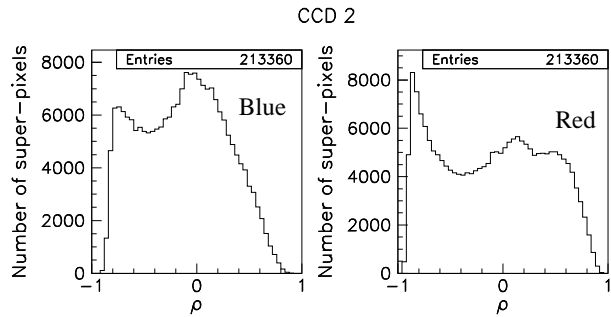}}
\caption{Histogram of the correlation coefficient \protect $\rho$
between the super-pixel flux and the seeing (before seeing
correction).}
\label{fig:corrsee}
\end{figure}
A correlation coefficient for each super-pixel $p$ can be defined
using the usual formula:
\begin{equation}
\rho^p = \frac{\sum_n \left(\phi^p_n - \phi \right) \left(S_n - S
\right)} {\sqrt{\sum_n \left(\phi^p_n - \phi \right)^2 \sum_n
\left(S_n - S\right)^2}}
\label{eq:corr}
\end{equation}
where $\phi$ and $S$ are the mean values of the super-pixel flux
$\phi^p_n$ and seeing $S_n$ on night $n$. In Fig. \ref{fig:corrsee},
we show the distributions of correlation coefficients $\rho^p$ in blue
(left) and in red (right) for each super-pixel $p$. These histograms
look quite different in both colours but both distributions have a
peak around $\rho \simeq -0.8$. This peak, which corresponds to the
anti-correlation with seeing near the centre of resolved stars, is
expected due to the large number of resolved stars. It is higher in
red than in blue, which is consistent with the EROS colour-magnitude
diagram where most detected stars have $B-R > 0$ (Renault
1996)\nocite{Renault:1996}. The correlation with seeing expected for
star tails ($\rho > 0$) is less apparent. However, a clear excess at
high values of $\rho$ (around $\rho \simeq 0.6$) appears in red, again
consistent with the EROS colour-magnitude diagram.  Figure
\ref{fig:clsee1} gives an example of such a correlation.  The upper
left panel of Fig. \ref{fig:clsee1} displays the scatter diagram of
one super-pixel flux versus the seeing, corresponding to a correlation
coefficient $\rho^p = -0.89$.  Despite the intrinsic dispersion of the
measurements (which could be large in particular when a temporal
variation occurs), a linear relationship is observed. The bottom left
panel displays the light curve of this super-pixel.

\subsection{Correction}
\begin{figure}
\resizebox{\hsize}{!}{\includegraphics{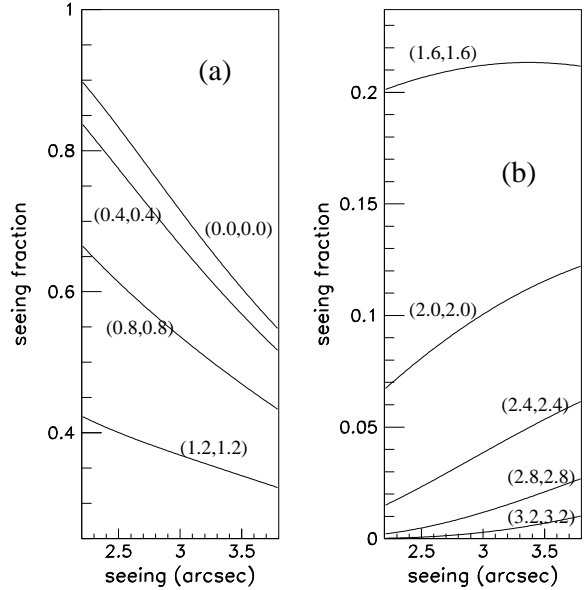}}
\caption{Variation of the seeing fraction of the star flux that enters
a $3\times3$ super-pixel as a function of seeing, for different star
positions, given on the figures in pixel unit with respect to the
centre of the super-pixel. Panel (a) shows cases when the centre of
the star lies within the super-pixel (anti-correlation). Panel (b)
shows the seeing fraction that could contribute from surrounding stars
(correlation).}
\label{fig:erf2}
\end{figure}
This seeing correction is aimed at eliminating the effect of the
seeing variations and to obtain pixel light curves that can be
described as the sum of a constant fraction of a centred star flux and
the background (see Eq.~\ref{eq:phipix}). The variation of the
super-pixel flux can be interpreted as a variation of the flux of this
centred stars (see Eq.~\ref{eq:delphipix}). However it is clear that
the super-pixel flux contains the flux of several stars and that we
are not doing stellar photometry, but rather super-pixel photometry.

The idea is to correct for the behaviour described in
Sect. \ref{sect:correl} using a linear expression:
\begin{equation}
{{\phi}^p_n}\vert_{\rm corrected} = \phi^p_n - \alpha^p \left({S_n -
S}\right) ,
\label{eq:see}
\end{equation}
where $\alpha^p$ is the estimate of the slope for each super-pixel and
${{\phi}^p_n}\vert_{\rm corrected}$ is the corrected flux. In the
following, $\phi^p_n$ will stand for this corrected flux.

Figure \ref{fig:erf2} shows that the seeing fraction of a given star
varies linearly with seeing, and hence justifies this correction.
\begin{figure}
\resizebox{\hsize}{!}{\includegraphics{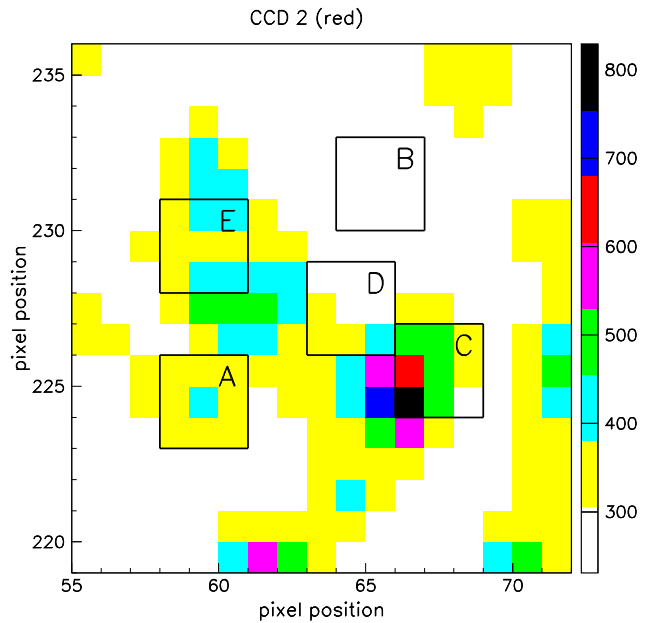}}
\caption{Different super-pixels labelled ``A'', ``B'', ``C''. ``D''
and ``E'' corresponding to the different configurations discussed in
the text.}
\label{fig:pat1}
\end{figure}
\begin{figure}
\resizebox{\hsize}{!}{\includegraphics{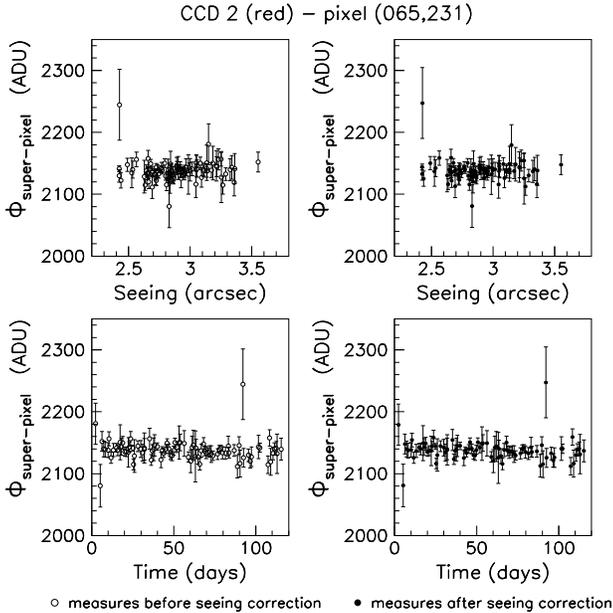}}
\caption{Seeing correction applied to the flux of a super-pixel light
curve with no significant correlation with the seeing \protect
($\rho^p =0.02$).  The position of the super-pixel on the CCD frame is
labelled ``B'' in Fig. \ref{fig:pat1}. The stars whose centres lie in
the super-pixel are too dim to be resolved.}
\label{fig:clsee2}
\end{figure}
\begin{figure}
\resizebox{\hsize}{!}{\includegraphics{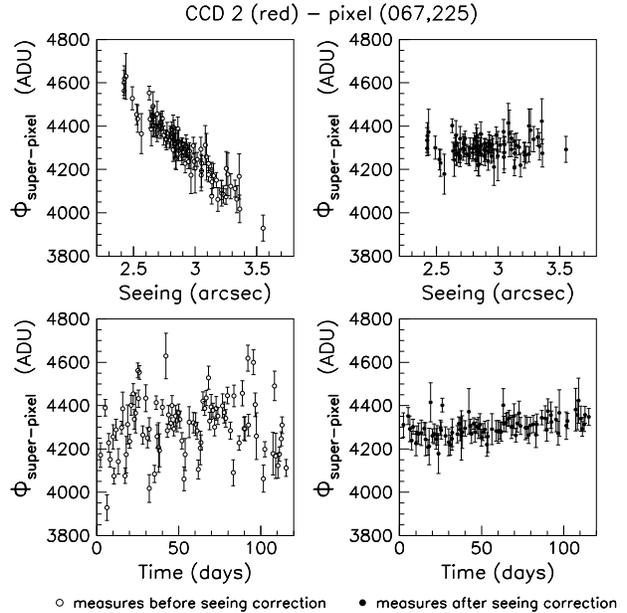}}
\caption{Seeing correction applied to the flux of a super-pixel light
curve strongly anti-correlated with the seeing \protect ($\rho^p
=-0.94$).  The position of the super-pixel on the CCD frame is
labelled ``C'' in Fig. \ref{fig:pat1}. This is a case of blending.}
\label{fig:clsee3}
\end{figure}
\begin{figure}
\resizebox{\hsize}{!}{\includegraphics{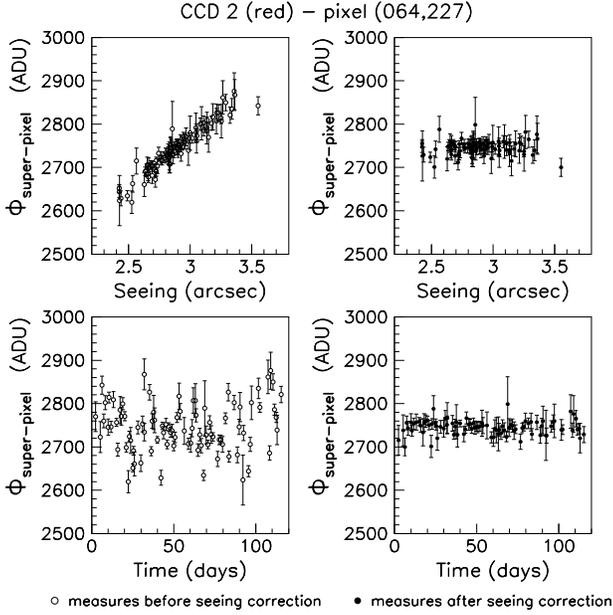}}
\caption{Seeing correction applied to the flux of a super-pixel light
curve strongly correlated with the seeing \protect ($\rho^p =0.96$).
The position of the super-pixel on the CCD frame is labelled ``D'' in
Fig. \ref{fig:pat1}. The super-pixel flux is dominated by tails of
surrounding stars. Centred stars are too dim to be resolved.}
\label{fig:clsee4}
\end{figure}
\begin{figure}
\resizebox{\hsize}{!}{\includegraphics{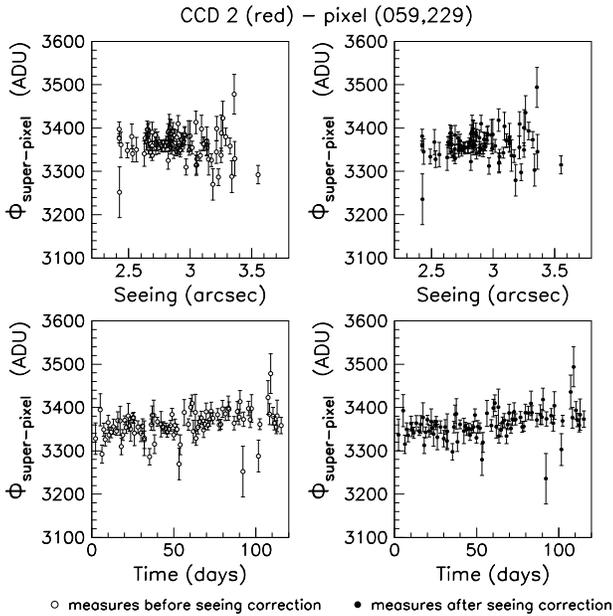}}
\caption{Seeing correction applied to the flux of a super-pixel light
curve not correlated with the seeing \protect ($\rho^p =0.05$).  The
position of the super-pixel on the CCD frame is labelled ``E'' in
Fig. \ref{fig:pat1}. Contributions due to the surrounding stars cancel
each other.}
\label{fig:clsee5}
\end{figure}
If several stars contribute to the super-pixel flux, their
contribution will add up linearly, because the flux of the background
$\phi_{bg}$ (Eq. \ref{eq:phipix}) can be written as:
\begin{equation}
\phi_{bg} = \sum_{i=1}^{N^{\rm other}_{\rm stars}} f_i \phi_i +
\phi^{\rm sky}_{\rm bg} ,
\end{equation}
where $i$ refers to the stars whose fluxes enter the super-pixel,
$f_i$ is the seeing fraction of each of these stars, $\phi_i$ their
flux, and $\phi^{\rm sky}_{bg}$ the sky background flux that enters
the super-pixel.  The first term describes the blending and crowding
components that can affect the pixel.

Different configurations can occur as shown in Fig. \ref{fig:pat1},
and are discussed in the following. Firstly, if there is no
significant contamination by surrounding stars, either (A in
Fig. \ref{fig:pat1}) the star flux is large compared to the noise that
affects the super-pixel or not (B in Fig. \ref{fig:pat1}). The effect
of the seeing correction on super-pixels of type A and B is shown in
Fig. \ref{fig:clsee1} and Fig. \ref{fig:clsee2}
respectively. Secondly, if there is a significant contamination by
surrounding stars, three cases must be considered:\vspace{-0.2cm}
\begin{itemize}
\item The centres of the surrounding stars lie in the super-pixel (C
in Fig. \ref{fig:pat1}; seeing correction in Fig. \ref{fig:clsee3}). 
\item The flux due to PSF wings of surrounding stars is larger than
the contribution of the centred star we are interested in (D in
Fig. \ref{fig:pat1}; seeing correction in Fig. \ref{fig:clsee4}). 
\item The flux due to PSF wings is comparable with the centred star
and their variation with the seeing cancel each other (E in
Fig. \ref{fig:pat1}; seeing correction in Fig. \ref{fig:clsee5}).
\end{itemize}

\subsection{Importance of the seeing correction}
\begin{figure}
\resizebox{\hsize}{!}{\includegraphics{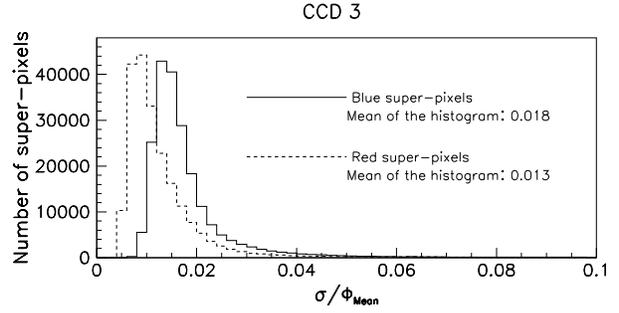}}
\caption{Relative flux stability achieved on {\it super-pixel} light
curves {\it after seeing correction} for all the pixels of CCD~3.}
\label{fig:stabi2}
\end{figure}
\begin{figure}
\resizebox{\hsize}{!}{\includegraphics{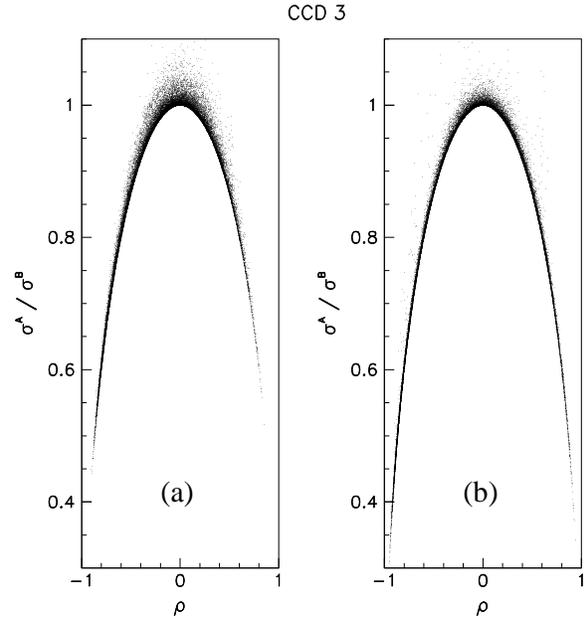}}
\caption{Importance of the seeing correction. The ratio
$\sigma^A/\sigma^B$ is displayed as a function of the correlation
coefficient $\rho$ calculated before the seeing correction is applied,
in blue (a) and in red (b). Data for $20\,000$ super-pixels are used.}
\label{fig:parabolla}
\end{figure}
The correction described above significantly reduces the fluctuations
due to seeing variations. Figure \ref{fig:stabi2} displays the
relative dispersion computed after this correction. With respect to
the histograms presented in Fig. \ref{fig:stabi1}, this dispersion is
reduced by 20\% in blue and 10\% in red, achieving a stability of
1.8\% in blue and 1.3\% in red, respectively 1.6 times the photon
noise in blue and 1.9 in red.  The improvement on the overall relative
stability remains modest, because most light curves {\it do not} show
a correlation with the seeing and {\it do not} need a correction.  The
importance of the seeing correction as a function of the correlation
coefficient $\rho$ can be more precisely quantified. Figure
\ref{fig:parabolla} displays for both colours the ratio $\sigma^A /
\sigma^B$, where $\sigma^A$ is the dispersion measured along the
super-pixel light curves {\em after} the seeing correction, and
$\sigma^B$ the one measured {\em before} the correction, as a function
of the initial correlation coefficient $\rho$. It can be shown that,
if the slope $\alpha$ defined in Eq. \ref{eq:see} is measured with an
error $\Delta \alpha$, then the following correlation is expected:
\begin{eqnarray}
\left({\frac{\sigma_A}{\sigma_B}}\right)^2 = {1-\rho^2} + {\Delta
\alpha
}^2 \left(\frac{\sigma_S}{\sigma_B}\right)^2 \nonumber
\end{eqnarray}
where ${\sigma_S}$ is the dispersion of the seeing.  This correlation
shows that the stronger the correlation with seeing, the more
important the seeing correction is.  The dispersion of the
measurements can be reduced up to 40\% for very correlated light
curves. The limitation of this correction comes from the errors
$\Delta a$ which explain why most points are slightly above this
envelope.  When $\vert \rho \vert < 0.15$, most points in fact lie
above 1, in which case the "correction" worthens things. Therefore we
do not apply the correction to light curves with $\vert \rho \vert <
0.15$.

As the seeing is randomly distributed in time, the above correction
will not induce artificial variations that could be mistaken for a
microlensing event or a variable star.

One can wonder however what happens to the super-pixel flux when the
flux of the contributing star varies.  In this case, the slope $a$ of
the correlation between the flux and the seeing does change, thus
resulting in a lower correlation coefficient. In extreme cases, when
the correction coefficient is small ($\vert \rho \vert < 0.15$), the
correction is thus not appropriate and not applied.

\subsection{Residual systematic effects}\hspace{0.5cm}
\label{sect:resi}
The seeing correction is empirical, and can be sensitive to bad seeing
determination due to inhomogeneous seeing across the image or a
(slightly) elongated PSF. Part of these problems is certainly due to
the atmospheric dispersion, as mentioned by Tomaney \& Crotts
(1996)\nocite{Tomaney:1996}. This phenomenon correlates with air mass,
and affects stars with different colour differently. This is a serious
problem for pixel monitoring as we do not know the colours of
unresolved stars.
\begin{figure}
\resizebox{\hsize}{!}{\includegraphics{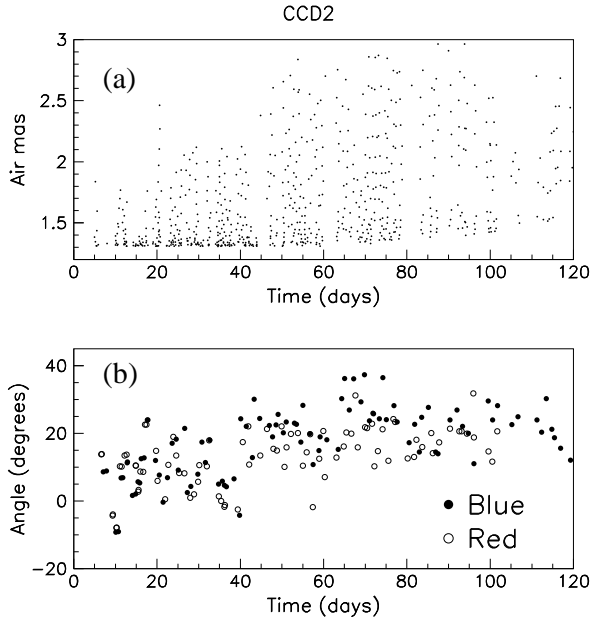}}
\caption{Possible systematics. Panel (a) displays the variations of
the air mass towards the LMC for each individual images. Panel (b)
shows the small-amplitude variations of the angle of rotation of the
PSF measured on the composite images.}
\label{fig:angair}
\end{figure}
Figure \ref{fig:angair} (a) displays the air mass towards the LMC as a
function of time for the images studied (before the averaging
procedure), and shows, besides a quite large dispersion of air mass
during the night, a slow increase with time. All the measurements have
an air mass larger than $1.3$, and half of them have an air mass
larger than $1.6$, producing non negligible atmospheric prism effects
because of the large passband of the filters. According to Filippenko
(1982)\nocite{Filippenko:1982}, photons at the extreme wavelengths of
our filters would spread over $0.73$ to $2.75$ arc-seconds in blue
depending on the air mass, and over $0.34$ to $1.17$ arc-second in
red.

While the PSF can be well approximated by a Gaussian (residuals
$\simeq$ 3\%), a more careful study shows that the PSF is elongated
with $\langle \sigma_b / \sigma_a \rangle \simeq 0.7$, where
$\sigma_b$ and $\sigma_a$ are the dispersions along the minor and
major axis of the ellipse.  However the fact that the PSF is elongated
does not affect the efficacy of the seeing correction: on the one
hand, for the central part of the stars, a similar seeing fraction
enters the super-pixel for a given seeing value; whereas on the other
hand, for pixels dominated by the tails of neighbouring stars, the
correlation of the flux with seeing will be slightly different, but
the principle remains the same. As the PSF function rotates up to
20$^\circ$ during the period of observation (see Fig.~\ref{fig:angair}
(b)), this could affect the super-pixels whose content is dominated by
the tail of one star and could produce spurious variations correlated
with the angle of rotation. Fortunately, this rotation is small and we
estimate that even in this unfavourable case it cannot produce
fluctuations of the super-pixel flux larger than 3\%, which can be
disturbing when close to bright stars. We expect this will produce the
kind of trends that can be observed in the bottom right panel of
Fig. \ref{fig:clsee3} and \ref{fig:clsee5}. {\it However this cannot
mimic any microlensing-like variation.}

We reach a level of stability close to photon noise, and this
stability can be expressed in terms of detectable changes in
magnitude: taking into account a typical seeing fraction $f=0.8$ for a
super-pixel, and assuming a total background characterised by a
surface magnitude $\mu_B \simeq 20$ in blue and $\mu_R \simeq 19$ in
red, stellar variability will be detected 5 $\sigma$ above the noise
if the star magnitude gets brighter than $20$ in blue and $19$ in red
at maximum.  With the Pixel Method, our ability to detect a luminosity
variation is not hindered by star crowding as we do not require to
resolve the star, whereas for star monitoring, the sample of monitored
stars is far from complete down to magnitude 20. Although the
dispersion measured along the light curves gives a good estimate of
the overall stability, we can refine it further and provide an error
bar for each super-pixel flux.

\section{Error estimates for super-pixel fluxes}
\label{section:opteb}\label{section:error}
As explained in previous sections, it is not straightforward to trace
the errors affecting pixel fluxes through the various
corrections. Errors are estimated here in a global way for each pixel
flux, ``global'' meaning that we do not separate the various sources
of noise.  The images used for the averaging procedure provide a first
estimate of these errors. The dispersion of the flux measurements
performed over each night allows the computation of an error
associated with the averaged pixel flux. We discuss how this estimate
deviates from Gaussian behaviour, and which correction can be
applied. Gaussian behaviour is of course an ideal case, but it
provides a good reference for the different estimates discussed here.

\paragraph*{Error estimates on elementary pixel.}\hspace{0.5cm}
When we perform for each night $n$ the averaging of pixel fluxes, we
also measure a standard deviation for each pixel
${\sigma_{\phi^p_n}}$. Assuming this dispersion is a good estimate of
the error associated to each flux measurement $\phi^p_{n,j}$, and that
the errors affecting each measurement are independent, we can deduce
an error ${\sigma^p_{n}}$ on $\phi^p_n$ as:
\begin{equation}
{\sigma^p_{n}}^2 = \frac{1}{N^p_n} {\sigma_{\phi^p_n}}^2
\label{eq:nois}
\end{equation}
This estimation, however, is uncertain: the number of images per night
can be quite small, and Eq. \ref{eq:nois} assumes identical weight for
all images of the same night.
\begin{figure}
\resizebox{\hsize}{!}{\includegraphics{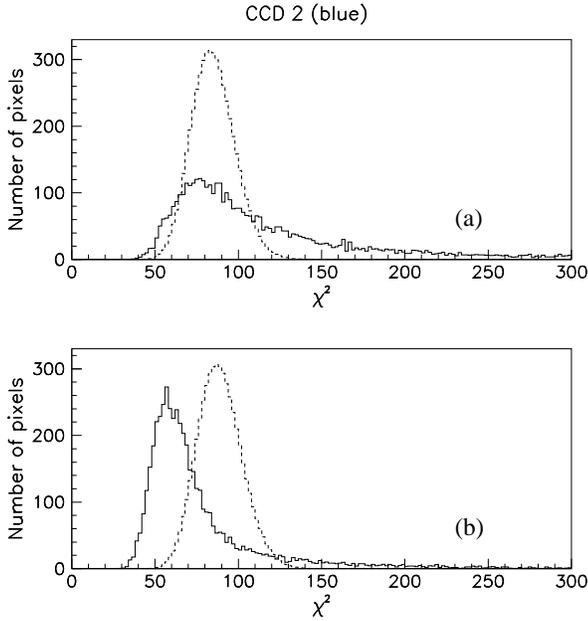}}
\caption{Distributions of ${\chi^p}^2$: on both panels, the dashed
line represents the ideal distribution discussed in the text.  Panel
(a) (solid line) displays the ${\chi^p}^2$ distribution with error
estimates based on the dispersion of pixel flux measurements over each
night. In panel (b), the histogram (solid line) is computed with
errors calculated for each pixel flux as the maximum between the
photon noise and the errors used in panel (a).}
\label{fig:chi2}
\end{figure}

\begin{figure}
\resizebox{\hsize}{!}{\includegraphics{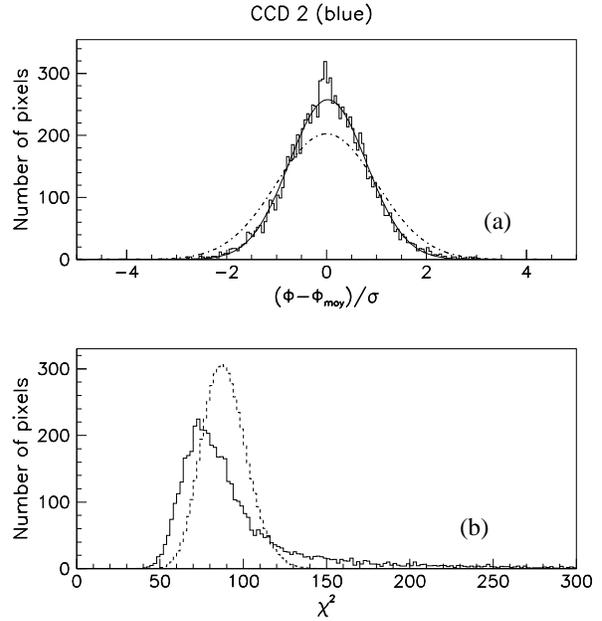}}
\caption{Corrected error bars: the upper panel (a) displays a \protect $z_n$
distribution for a given image \protect $n$ whose errors were
over-estimated (histogram). The full line corresponds to the fitted
Gaussian distribution, and the dashed line to the normalised Gaussian
distribution. The lower panel (b) shows the \protect $\chi^2$
distribution calculated with the corrected errors (solid line).}
\label{fig:gauss}
\end{figure}

In order to assess our error estimates, we compute the distribution of
the ${\chi^p}^2$ values associated with each pixel $p$ light curve.
\begin{equation}
{\chi^p}^2 = \sum_n\frac{\left({\phi^p_n - \langle \phi^p
\rangle}\right)^2} {{\sigma^p_n}^2}
\end{equation}
Figure \ref{fig:chi2}(a) displays two $\chi^2$ distributions: the
ideal case (dashed line) assumes Gaussian noise and the number of
degree of freedom (hereafter NDOF) of the data\footnote{Since only one
image is available for 3 of the nights, the corresponding points do
not have any error bars at this stage, but will have one in the next
one. This explains why the ideal Gaussian distribution of
Fig. \protect\ref{fig:chi2}(a) (dashed line) is slightly shifted
towards the left with respect to those in Fig. \ref{fig:chi2}(b) and
Fig. \ref{fig:gauss}(b).}; the solid line uses actual data with errors
computed with Eq. (\ref{eq:nois}): the histogram peaks roughly to the
correct NDOF, but exhibits a heavy tail corresponding to non-Gaussian
and under-es\-ti\-ma\-ted errors.

Due to statistical uncertainties on the calculation of the errors
$\sigma^p_n$, it happens that some of them are estimated to be smaller
than the corresponding photon noise, in which case the photon noise is
adopted as the error.  The corresponding $\chi^2$ distribution
displayed (solid line) in Fig.  \ref{fig:chi2}(b) has a smaller
non-Gaussian tail, but peaks at a smaller NDOF: not surprisingly the
errors are now over-estimated.

\paragraph*{Correction.}\hspace{0.5cm}
The correction described here is intended to account for {\it night to
night variations}, or important systematic effects altering some
images. Although the main variations in the observing conditions have
been eliminated by the procedures described above, each night is
different and for instance the seeing distribution over one night can
differ from the global one.  Hence, we weight each error with a
coefficient depending on the composite image. The principle is to
consider the distribution for each night $n$ of the variable $z_n^p$
given by
\begin{equation}
z_n^p = \frac{{\phi^{p}_{n}} -
{\langle{{\phi^{p}}}\rangle}_n}{\sigma^p_n} ,
\end{equation}
and to re-normalise it in order to approach a normal Gaussian
distribution as well as possible.  ${\langle{{\phi^{p}}}\rangle}_n$ is
the mean pixel $p$ flux value computed over the whole light curve.
The standard deviation $\sigma^\prime_n$ of each of these $z_n^p$
distributions is estimated for each average image $n$ on a central
patch\footnote{The values of this dispersion $\sigma^\prime_n$
fluctuate around 4\% from patch to patch.} of 100$\times$100 pixels.
A $z_n^p$ distribution is plotted for each image and is fitted with a
Gaussian distribution.  This fit is quite good for most of the images
and the dispersion of the Gaussian distribution is our estimate of
$\sigma^\prime_n$.  Figure \ref{fig:gauss}(a) shows an example of the
$\sigma^\prime_n$ estimation. The solid line shows a Gaussian fit to
the data. The width is not equal to $1$ as it should be, but rather to
$0.77$, the value of $\sigma^\prime_n$ for this image. For comparison,
we show a Gaussian of width $1$, with the same normalisation (dashed
line).

In the following, the corrected errors
\begin{equation}
{{\sigma}^p_n}\vert_{\rm corrected} = \sigma^\prime_n {{\sigma}^p_n} ,
\end{equation}
are associated with each pixel flux. $\sigma^p_n$ is different for
each measurement whereas $\sigma^\prime_n$ is a constant for each
image $n$. The resulting $\chi^2$ histogram is displayed in
Fig. \ref{fig:gauss}(b) (full line). The $\chi^2$ distribution peaks at
a higher value of $\chi^2$ than before correction
(Fig. \ref{fig:chi2}(b)), which however is still slightly smaller than
the NDOF.

\paragraph*{From pixel errors to super-pixel errors.}\hspace{0.5cm}
We have seen in Sect. \ref{subsection:stabi} that the use of
super-pixel light curves allows us to reduce significantly the flux
dispersion along the light curves. The most natural approximation for
the computation of super-pixel errors is to assume those on elementary
pixels to be independent:
\begin{equation}
{\sigma^{sp}_n} = \sqrt{\sum_{p} {{\sigma}^p_n}^2} .
\end{equation}
However, errors on neighbouring pixels are not independent, because of
the geometrical alignment procedure and of the seeing correction. To
take this into account, we correct the error on super-pixels in the
same way as above. The factors $\sigma^\prime_n|_{sp}$ thus obtained
are 20\% higher than for elementary pixels.
\begin{figure}
\resizebox{\hsize}{!}{\includegraphics{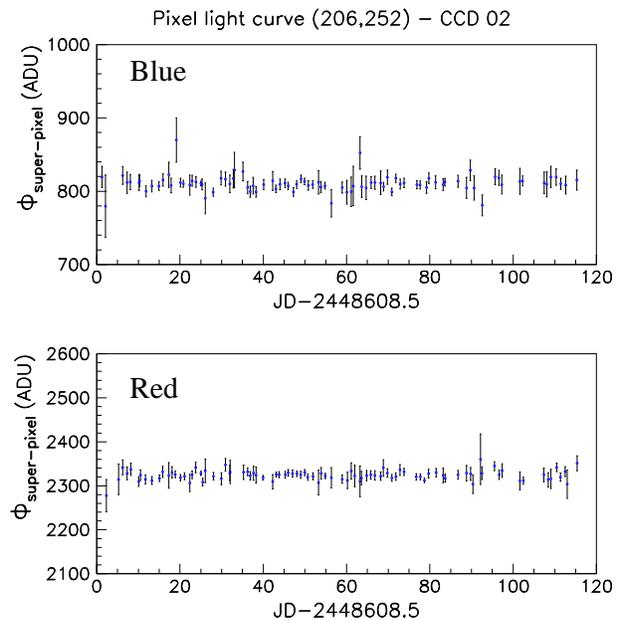}}
\caption{Example of a stable super-pixel light curve.}
\label{fig:CLstable}
\end{figure}
\begin{figure}
\resizebox{\hsize}{!}{\includegraphics{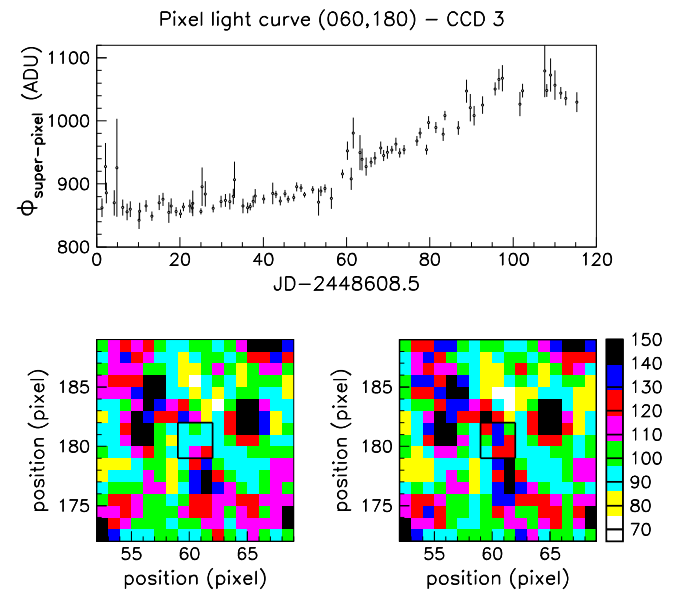}}
\caption{Example of a variable super-pixel light curve in blue
(top). The star is unresolved at minimum (bottom left panel) and would
even be difficult to detect with classical procedures at maximum
(bottom right panel).}
\label{fig:clvar}
\end{figure}

We have now super-pixel light curves with an error estimate for each
flux. Figure \ref{fig:CLstable} displays an example of a typical
stable light curve in blue (upper panel) and in red (lower panel),
whereas Fig. \ref{fig:clvar} is an example of a variable light
curve. The error bars obtained as a combination of expected photon
noise and experimental data are compatible with the dispersion
observed on the super-pixel light curves, and allow us to give a
different weight to each measurement according to its quality.

\section{Conclusion}
\label{section:simu}
The treatment described here has produced $2.1 \times 10^6$
super-pixel light curves corrected for observational variations, with
an error bar for each point. They are characterised by an average
stability close to twice the photon noise: dispersions of 1.8\% of the
flux in blue and 1.3\% in red are measured over a 120 days time
period.  To reduce the effects of the dispersion due to the
observational conditions, averaging the images of each night turns out
to be a crucial step.  The fluctuations due to seeing variations have
been corrected for.  We associate an error bar with each measurement, and
these careful estimates together with the study of possible
systematics are used in the companion papers for the detection of
intrinsic luminosity variations.

This study is the starting point for the comprehensive microlensing
search described in Paper II. The error estimates enter the definition
of the selection criteria and constitute an important ingredient for
microlensing Monte-Carlo simulations required to quantify the
efficiency of the pixel microlensing method.  The study of the
background of variable stars will be addressed in Paper III.

\begin{acknowledgements}
We thank A. Gould for extremely useful discussions and suggestions,
and D. Valls-Gabaud for a careful reading of the manuscript.  We are
also particularly grateful to Claude Lamy for her useful help on data
handling during this work. P. Gondolo was partially supported by the
European Community (EC contract no. CHRX-CT93-0120).  A.L. Melchior
has been supported by grants from the Singer-Polignac Foundation, the
British Council, the DOE and by NASA grant NAG5--2788 at Fermilab.
\end{acknowledgements}


\begin{thebibliography}{}

\bibitem[\protect\astroncite{Alard et~al.}{1995}]{Alard:1995}
Alard, C., et~al.: 1995,
\newblock {\em  ESO Messenger}, {\bf 80}, 31

\bibitem[\protect\astroncite{Alcock et~al.}{1993}]{Alcock:1993}
Alcock, C., et~al.: 1993, \newblock {\em Nat} {\bf 365}, 621

\bibitem[\protect\astroncite{Alcock et~al.}{1995}]{Alcock:1995a}
Alcock, C., et al.: 1995,
\newblock {\em ApJ} {\bf 445}, 133

\bibitem[\protect\astroncite{Alcock et~al.}{1996}]{Alcock:1996}
Alcock, C., et al.: 1996,
\newblock {\em ApJ} {\bf 486}, 697

\bibitem[\protect\astroncite{Alcock et~al.}{1998}]{Alcock:1998}
Alcock, C., et al.: 1998,
\newblock {\em ApJ} {\bf 499}, L9

\bibitem[\protect\astroncite{Ansari}{1994}]{Ansari:1994}
Ansari, R.: 1994,
\newblock {\em Une m\'ethode de reconstruction photom\'etrique pour
  l'exp\'erience EROS},
\newblock preprint LAL 94-10

\bibitem[\protect\astroncite{Ansari et~al.}{1997}]{Ansari:1997a}
Ansari, R., et al., G.: 1997,
\newblock {\em A$\&$A} {\bf 324}, 843

\bibitem[\protect\astroncite{Arnaud et~al.}{1994a}]{Arnaud:1994b}
Arnaud, M. et~al.: 1994a,
\newblock {\em Experimental astronomy} {\bf 4}, 265

\bibitem[\protect\astroncite{Arnaud et~al.}{1994b}]{Arnaud:1994a}
Arnaud, M. et~al.: 1994b,
\newblock {\em Experimental astronomy} {\bf 4}, 279

\bibitem[\protect\astroncite{Ashman}{1992}]{Ashman:1992}
Ashman, K.~M.: 1992, \newblock {\em PASP} {\bf 104}, 682

\bibitem[\protect\astroncite{Aubourg et~al.}{1993}]{Aubourg:1993}
Aubourg, E., et al.: 1993, \newblock {\em Nat} {\bf 365}, 623

\bibitem[\protect\astroncite{Aubourg et~al.}{1995}]{Aubourg:1995}
Aubourg, E., et al.: 1995, \newblock {\em A$\&$A}
{\bf 301}, 1A

\bibitem[\protect\astroncite{Baillon et~al.}{1993}]{Baillon:1993}
Baillon, P., et al.: 1993, \newblock {\em A$\&$A} {\bf 277}, 1

\bibitem[\protect\astroncite{Carollo et~al.}{1995}]{Carollo:1995}
Carollo, C.~M., et al.: 1995, \newblock {\em ApJ} {\bf 441}, L25

\bibitem[\protect\astroncite{Carr}{1994}]{Carr:1994}
Carr, B.: 1994, \newblock {\em ARA$\&$A} {\bf 32}, 531

\bibitem[\protect\astroncite{Copi et~al.}{1995}]{Copi:1995}
Copi, C.~J., et al.: 1995, \newblock {\em Sci} {\bf 267}, 192

\bibitem[\protect\astroncite{Crotts}{1992}]{Crotts:1992}
Crotts, A. P.~S.: 1992, \newblock {\em ApJ} {\bf 399}, L43

\bibitem[\protect\astroncite{Evans}{1994}]{Evans:1994}
Evans, N.~W.: 1994, \newblock {\em ApJ} {\bf 437}, L31

\bibitem[\protect\astroncite{Filippenko}{1982}]{Filippenko:1982}
Filippenko, A.~V.: 1982, \newblock {\em PASP} {\bf 94}, 715

\bibitem[\protect\astroncite{Gerhard and Silk}{1996}]{Gerhard:1996}
Gerhard, O. and Silk, J.: 1996, \newblock {\em ApJ} {\bf 472}, 34

\bibitem[\protect\astroncite{Giraud-H\'eraud}{1997}]{YGH:1997}
Giraud-H\'eraud, Y.: 1997, \newblock {\em AGAPE, Andromeda
Gravitational And Pixel Experiment}, \newblock 3rd Microlensing
Workshop, Notre-Dame, USA

\bibitem[\protect\astroncite{Gould}{1995}]{Gould:1995b}
Gould, A.: 1995, \newblock {\em ApJ} {\bf 455}, 44G

\bibitem[\protect\astroncite{Gould}{1996a}]{Gould:1996b}
Gould, A.: 1996a, \newblock in {\em Sheffield workshop on
Identification of Dark Matter - astro-ph/9611185}

\bibitem[\protect\astroncite{Gould}{1996b}]{Gould:1996}
Gould, A.: 1996b, \newblock {\em ApJ} {\bf 470}, 201

\bibitem[\protect\astroncite{Grison et~al.}{1995}]{Grison:1994b}
Grison, Ph., et~al.: 1995, \newblock {\em A$\&$AS} {\bf 109}, 447

\bibitem[\protect\astroncite{Kerins}{1996}]{Kerins:1997}
Kerins, E.~J.: 1997, \newblock {\em A$\&$A} {\bf 322}, 709

\bibitem[\protect\astroncite{Melchior}{1998a}]{Melchior:1998a}
Melchior, A.-L. et~al.: 1998a, \newblock {\em A$\&$A submitted,
Paper~II}	
	
\bibitem[\protect\astroncite{Melchior}{1998b}]{Melchior:1998b}
Melchior, A.-L. et~al.: 1998b, \newblock {\em in preparation,
Paper~III}

\bibitem[\protect\astroncite{Paczynski}{1986}]{Paczynski:1986}
Paczy\'nski, B.: 1986, \newblock {\em ApJ} {\bf 304}, 1

\bibitem[\protect\astroncite{Persic and Salucci}{1992}]{Persic:1992}
Persic, M. and Salucci, P.: 1992, \newblock {\em MNRAS} {\bf 258}, 14P

\bibitem[\protect\astroncite{Queinnec}{1994}]{Queinnec:1994}
Queinnec, F.: 1994, \newblock {\em Ph.D. thesis}, Universit\'e de
Paris VII, Paris

\bibitem[\protect\astroncite{Renault}{1996}]{Renault:1996}
Renault, C.: 1996, \newblock {\em Ph.D. thesis}, Universit\'e de Paris
VII, Paris

\bibitem[\protect\astroncite{Stanek et~al.}{1997}]{Stanek:1997}
Stanek, K.~Z. et~al.: 1997, \newblock {\em ApJ} {\bf 477}, 163

\bibitem[\protect\astroncite{Tomaney and Crotts}{1996}]{Tomaney:1996}
Tomaney, A.~B. and Crotts, A.: 1996, \newblock {\em AJ} {\bf 112}, 2872

\bibitem[\protect\astroncite{Udalski et~al.}{1995}]{Udalski:1995b}
Udalski, A., et al.: 1995, \newblock {\em Acta Astron.} {\bf 45}, 237

\bibitem[\protect\astroncite{Walker et~al.}{1991}]{Walker:1991}
Walker, T.~P., et al.: 1991, \newblock {\em ApJ} {\bf 376}, 51

\bibitem[\protect\astroncite{White et~al.}{1996}]{White:1996}
White, M., et al.: 1996, \newblock {\em MNRAS} {\bf 283}, 107

\bibitem[\protect\astroncite{Zaritsky}{1992}]{Zaritsky:1992}
Zaritsky, D.: 1992, \newblock {\em PASP} {\bf 104}, 831

\end{thebibliography}
\end{document}